# SlopMap: a software application tool for quick and flexible identification of similar sequences using exact k-mer matching

Ilya Y. Zhbannikov[*1], Samuel S. Hunter[†1,2], Matthew L. Settles[‡1,2], and James A. Foster[§1,2,3]

[1]Bioinformatics and Computational Biology Program, University of Idaho, USA
[2]The Institute of Bioinformatics and Evolutionary Studies (IBEST), University of Idaho, USA
[3]Department of Biological Sciences, University of Idaho, USA

## Abstract

With the advent of Next-Generation (NG) sequencing, it has become possible to sequence an entire genome quickly and inexpensively. However, in some experiments one only needs to extract and assembly a portion of the sequence reads, for example when performing transcriptome studies, sequencing mitochondrial genomes, or characterizing exomes. With the raw DNA-library of a complete genome it would appear to be a trivial problem to identify reads of interest. But it is not always easy to incorporate well-known tools such as BLAST, BLAT, Bowtie, and SOAP directly into a bioinformatics pipelines before the assembly stage, either due to incompatibility with the assembler's file inputs, or because it is desirable to incorporate information that must be extracted separately. For example, in order to incorporate flowgrams from a Roche 454 sequencer into the Newbler assembler it is necessary to first extract them from the original SFF files.

We present SlopMap, a bioinformatics software utility which allows rapid identification similar to provided target sequences from either Roche 454 or Illumnia DNA library. With a simple and intuitive command-line interface along with file output formats compatible with assembly programs, SlopMap can be directly embedded in biological data processing pipeline without any additional programming work. In addition, SlopMap preserves flowgram information needed for Roche 454 assembler.

[*] Electronic address: zhba3458@vandals.uidaho.edu; Corresponding author
[†] Electronic address: shunter@uidaho.edu
[‡] Electronic address: msettles@uidaho.edu
[§] Electronic address: foster@uidaho.edu

# 1 Introduction

New methodologies enabled by Next Generation Sequencing (NGS) that are of particular interest to us include transcriptome analysis for RNA research [4] and mitohondrial sequencing from exome data [3]. Such applications include those in which the researcher is interested in assembling only specific content within a genome of interest, using a set of targets to initialize the assembly process. It may seem trivial to identify the reads of interest among those produced by NGS hardware, using well-known general-purpose alignment or mapping tools such as Blat [7], Bowtie2 [8], BWA [9], and SOAP [5]. But even an efficient tool may be difficult to incorporate directly into a bioinformatics pipeline before the assembly stage, since it may be necessary to convert data to a different file format. For example, existing mappers usually use the SAM/BAM [6] file format as output. None use SFF format files [1] as both input and output, and none but Bowtie2 support FastQ output, and it is supported only in a limited sense.

Moreover, it is difficult to use existing mapping software tools when it necessary to establish a similarity threshold, i.e. when one wants reads that are 50%, 70% or 85% similar to the target (Figure 1). Relying only on input parameters such as gap penalties and seed size, which most well-known aligners have, it is difficult to achieve flexible mappings with require percentage of similarity. On the other hand, it it often desirable to find reads that are at least 90% similar to the provided target, and to discard the rest.

Another problem arises if there is insufficient data on the edge of the target located within a reference genome (Figure 1). In this situation the whole read (marked red) can potentially be discarded due to lack of data on the edge, even if a part of the read has significant similarity to the target.

We present SlopMap, a bioinformatic software utility that quickly and flexibly identifies sequence reads that are within a given percent similarity to a target sequence. SlopMap is not a sequence alignment mapper, but rather identifies reads which may have been derived from the target region. Unlike traditional alignment software, SlopMap only reports reads that are similar to the provided target. SlopMap selects reads for downstream analysis, such as assembly of sub-genome targets i.e. bacterial plasmids, virae, mitochondria, exome capture data, chloroplasts, transcriptomes, etc. It employs exact kmer matching, which we call sloppy mapping, without conducting the computationally expensive alignment stage of traditional mappers. SlopMap can be directly embedded in biological data processing pipelines before an assembly stage, since it maintains file format and preserves the original information such as bases, quality scores, and flowgrams (in the case of SFF files). SlopMap accepts both SFF (Roche 454 or Life Sciences Ion Torrent/Proton) and fastq (Illumina) file formats. SlopMap is a simple, easy to use and robust tool that can be used with percent similarity to targets as low as 5% (95% dissimilar).

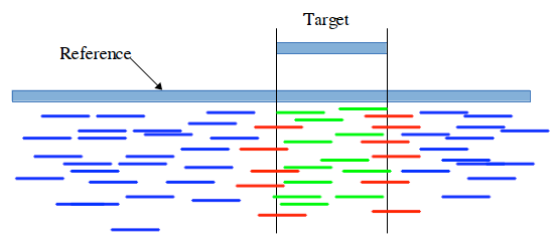

Figure 1: Situation when red reads on the edge of the target can be discarded by the general-purpose mapper. Green sequences can still be reported.

SlopMap along with its user manual is freely available under GPL from Bitbucket: http://bitbucket.org/izhbannikov/slopmap/.

## 2 Method

SlopMap is fully implemented and optimized in C++ for efficiency. This is a command-line application with all the input parameters specified on the command line. SlopMap is tunable via input parameters for kmer size k, percentage of similarity t and distance between two consecutive kmers d. It also supports flexible input and output file formats: FastQ, FastA, SFF and TXT.

### 2.1 Input Files

#### 2.1.1 Target

The target library file, also known as a "database", is a FastA formatted file that contains one or multiple records. Each record consists of two parts: a header and a sequence string. The header must contain a name which is a unique identifier of the record. The sequence string is DNA sequence which specifies the target of interest.

#### 2.1.2 DNA Query Library Files

The DNA query library files are data from the NGS machines. SlopMap can take either Roche 454 SFF or FastQ formatted files, or Illumina paired- or single-end reads. SlopMap computes the similarity of each read in a query library file to record in a target library.

### 2.2 Search procedure

SlopMap employs a multi-kmer search approach with single-base overlap (that can be increased) between two consecutive kmers in order to quickly determine the similarity to the target record. The algorithm is simple and straightforward:

1. Compute a dictionary from the given target library.

2. For each query sequence ("read") in the DNA query library:

    - Compute a set of consecuitive kmers.
    - Calculate the read's similarity to the set of target sequences.

3. Output summary statistics and files.

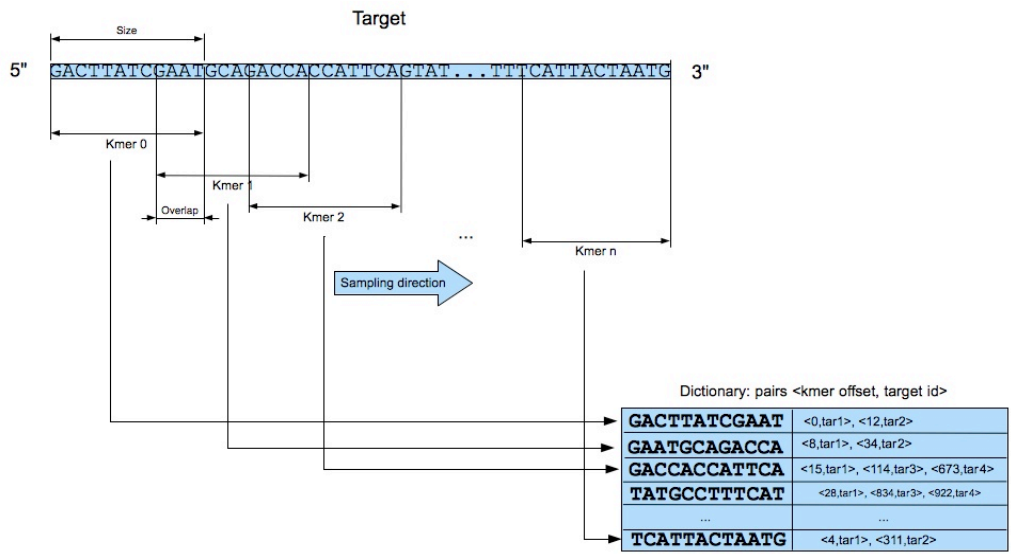

Figure 2: Sampling the target with constant pre-defined kmer size k and a distance d between two consecuitive kmers.

### 2.2.1 Dictionary

The first step is to build a dictionary indexed with kmers by sampling the given target (Figure 2). The kmers are short (usually 9-15 bases long) substrings representing the contiguous target. The target library sequences are sampled with the following pre-defined parameters: kmer size and a distance representing the constant overlap between consecutive kmers. These parameters remains unchanged throughout program execution. All kmers are hashed and associated with the offset position in the target string and a target record id. By default SlopMap uses kmer size of 15 bases. Google Dense Hash Map [2] that allows far fast data retrieval and memory efficiency is used as a data structure for the kmer dictionary.

### 2.2.2 Kmer matching

Query strings are sequentially sampled, so each query string contained within a given DNA query library is handled individually. Kmer size and a distance between two consecutive kmers remains the same for all reads in the query library. For each read taken from the query library, a dictionary search is performed and similarity between the read and the target is calculated as follows:

$$S = \frac{\text{number of shared kmers} * \text{kmerLength}}{Min\_length(\text{query string}, \text{target string})}, \quad 0 \leq S \leq 1$$

Where number shared bases is the total length of kmers shared between the target and query string, Min_length(query string, target string) is the minimum length, either of the query or target. Those reads that meet the pre-specified similarity threshold (parameter "−t" in SlopMap, by default it is set to 0.75) are then saved in output files, others are ignored. The values of S range from 0, having no similar kmers to the target, to 1, having all kmers shared between the read and target.

### 2.3 Output files

The output files are:

- A report file that contains all information about the reads, including similarity, positions of first and last match within the target, bases and quality scores.

- One or two FastQ formatted file(s), depending on whether the data are from single- or paired-end library.

- Optionally, an SFF formatted file, which contains only those Roche 454 sequences similar to the provided target.

## 3 Validation

To validate SlopMap we compared it to several alternative DNA mapping tools: Bowtie2, BWA, Blat, on two different query DNA libraries: 621,578 Roche 454 Esheria coli K12 W583 reads; and 3,875,453 Illumina Yeast Saccharomyces cerevisiae W303-K6001 reads. For the target sets, we randomly chose ten genes with various lengths from both genomes, each of which has over 4000 genes: thrA, thiQ, cydD, ycgB, dhaR, alkA, yfgF, yphE, mscS, parC (E-coli, GenBank accession number: U00096.2); YNL095C, YNL094W, YNL093W, YNL092W, YNL091W, YNL090W, YNL089C, YNL088W, YNL087W, YNL085W (Yeast, GenBank accession number: AF458977.1).

We estimate the number of reads found by SlopMap with various thresholds and kmers in order to:

- Estimate the effect of threshold to the number of reads found by SlopMap. Suggest values of t and k for optimal read recovery.

- Estimate the effect of various values of d (distance between two consequtive kmers) to the number of reads found by SlopMap against various threshold values allowing us to determine the range of optimal values for d.

- Compare the number of reads found by SlopMap to the number of reads found by other tools. In particular, to answer the question: what are the threshold values where read sets found by other tools are still subsets of reads found by SlopMap.

All tests were performed on a Linux server with Dual-Core AMD Opteron 8216 2.4 GHz processors (32 processors total) and 1 TB of shared memory and a laptop with single Intel Core i3 processor (four cores) with 4GB of memory.

## 4 Results and discussion

We calculated the number of found reads using various threshold value and kmer sizes and compared our results to existing read mappers. These results are presented in Figure 3, which shows the number of reads against various distances between two consecutive kmers. Further comparisons are made to compare the overlap between recovered read sets. These results are presented as Venn diagrams in Figures 5 and 6. We also provided recommendation for optimal values of parameters k and d.

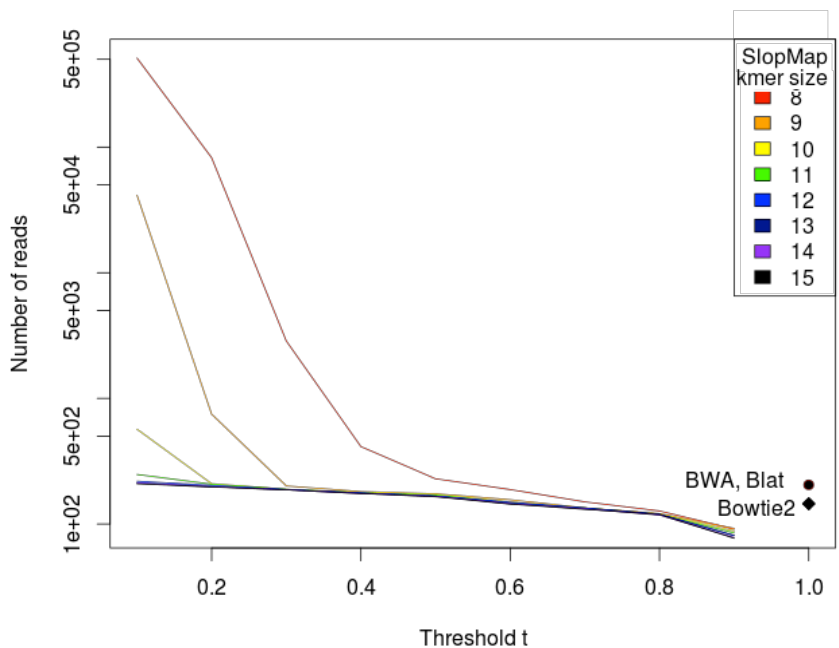

Figure 3: Number of reads found by SlopMap versus various percent identity threshold values t and different kmer lengths k. At the left side of the plot there is large amount of reads found for k ≤ 9 and t ≤ 0.2 showing that these parameter values may result in many false-positives. Larger values of k and t can be used to be more selective in read recruitment, or in situations where the reference is highly similar to the sequenced reads. Roche 454 E. coli K12 W583 DNA library (621,578 reads) was used as a query library for this test.

## 4.1 Number of reads found by SlopMap using different threshold values and kmer lengths

Figure 3 shows the number of reads found by SlopMap using different kmer sizes and threshold values. From this plot its easy to see that for kmer values k, of 7, 8, 9 along with threshold value less than 0.1 (10% of similarity), large amount of reads found. These values of k should be used only in such situations where the forged sequence is very divergent from the sequenced sample. Low values of t and k result in high sensitivity at the expense of specificity and should be used carefully to avoid multiple false positive hits.

Values of k within the range 10...15 can be used to generate more specific matches and are recommended for general usage. Higher values of k will result in less false positive mapping, SlopMap will not match kmers with mismatches, and will fail match reads at higher values of k. This is especially true in situations where reads have errors or adapters, which will generate false kmers and where there are real variants between the target and sequenced sample.

## 4.2 Number of reads found by SlopMap using various threshold values and distances between two consecutive kmers

In order to examine how different values of parameter d (the distance between two consecutive kmers in a read) impacts to the number of reads found by SlopMap and propose the optimal value for d, we provided a set of tests with kmer size k = 11 and threshold values t = 0.1...1.0 (i.e. from 10% to 100% similarity). This is shown in Figure 4. We find that in this data set, d has minimal impact on read recruitment. However we observe a higher recruitment rate for lower d, suggesting higher sensitivity. Using a small d results in slower performance however, so in cases where target and reference are highly divergent, a low d should be used, while a higher d can be used for more similar sequences. With these considerations in mind, we have set the default value of d to 3, and allow the user to change it using the command-line parameter −d = N. Our recommendations for the parameter d value to be no more than kmer length. Otherwise there may not be sufficient coverage. The values for the d from 1 to 5 are optimal for k = 11 bases, since they give number of reads significantly higher than other tools within t ≤ 0.5 (t = 0.75 is set by default in SlopMap).

## 4.3 Sensitivity test: comparison the number of reads reported by SlopMap and other tools

Gaps occur when part of the query aligns to one part of the reference and another part aligns close to the first part but with a gap of one or more bases. Such gaps are usually well recognized by some widely used aligners. Another type of gap can occur at the end of a target sequence, when part of the query matches the target, resulting in an end gap. SlopMap can find such reads and thereby identify more similar sequences than some other alignment tools. In order to compare the sensitivity of SlopMap to a set of other mappers (BWA, Blat, Bowtie2), we conducted several tests using Roche 454 reads from E.coli mapped against ycgB gene sequence. We are interested what is the cut-off point when reads found by alignment tools are still subsets of reads found by SlopMap. We can roughly say that the set of reads reported by one application is a subset of reads reported by another application if there is more 95% overlap between these two sets. For two kmer lengths (10 and 15 bases) and threshold values (0.1...1 with step of 0.1) we computed Venn diagrams that show overlap sets reported by SlopMap and other tools. From these diagrams we conclude that for threshold values below 0.3 (30% similarity) and for both kmer sizes (10 and 15), the reads found by Blat is a subset of number of reads found by SlopMap. Threshold values when reads reported by Bowtie 2 and BWA are still subsets of number of reads reported by SlopMap are 0.7 and 0.3 (70% and 30% similarity) for kmer size 10 and 15 respectively. Results are shown in Figures 5 and 6.

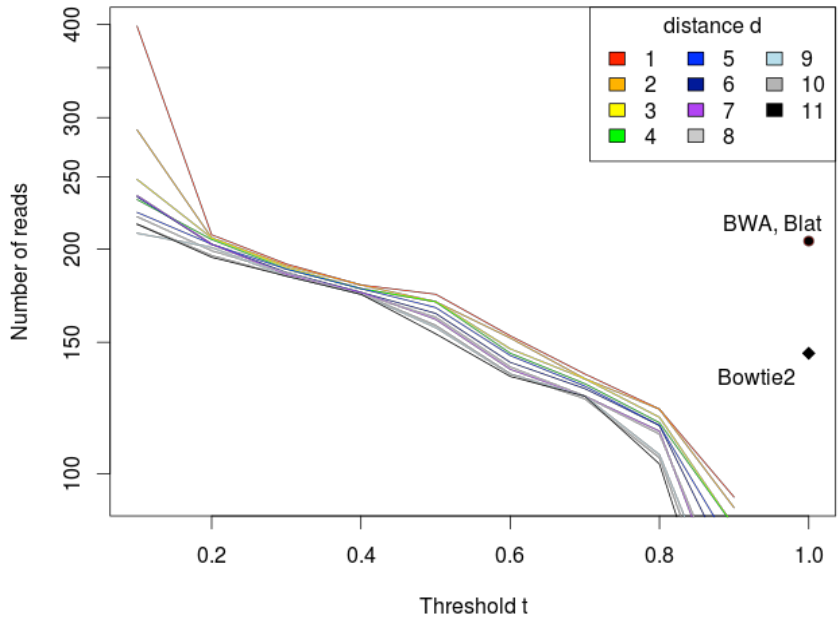

Figure 4: Number of reads found by SlopMap against various threshold values and different distances between two consecutive kmers. Roche 454 E. coli K12 W583 DNA library (621,578 reads) mapped against a single bacterial gene ycgB(1583 bp) was used as a query library for this test.

When threshold value t is 100%, SlopMap does not find any reads similar to the target sequence. This is expected because reads can contain base-call errors, homopolymers and other artefacts that introduce noise into sequences.

### 4.4 Non-consecutive matches

When we compute the similarity of reads we do not assume that kmer matches are consecutive. Non-consecutive matching may occur in situation which are biologically possible such as exon shuffling, inversion, etc. In Figure 7, the read and a target are shown along with kmers shared between them. In this situation, kmer 1, 2 and 3 match corresponding kmers in a target but in different order (non-consecutive). The read can be still considered as similar to the target. SlopMap identifies and reports this read as being similar to the target, despite the rearrangement.

### 4.5 Timing considerations

Figure 8 displays the execution time required to complete each search. We compare execution times for various threshold values of SlopMap (other parameters were set to default) with other tools.

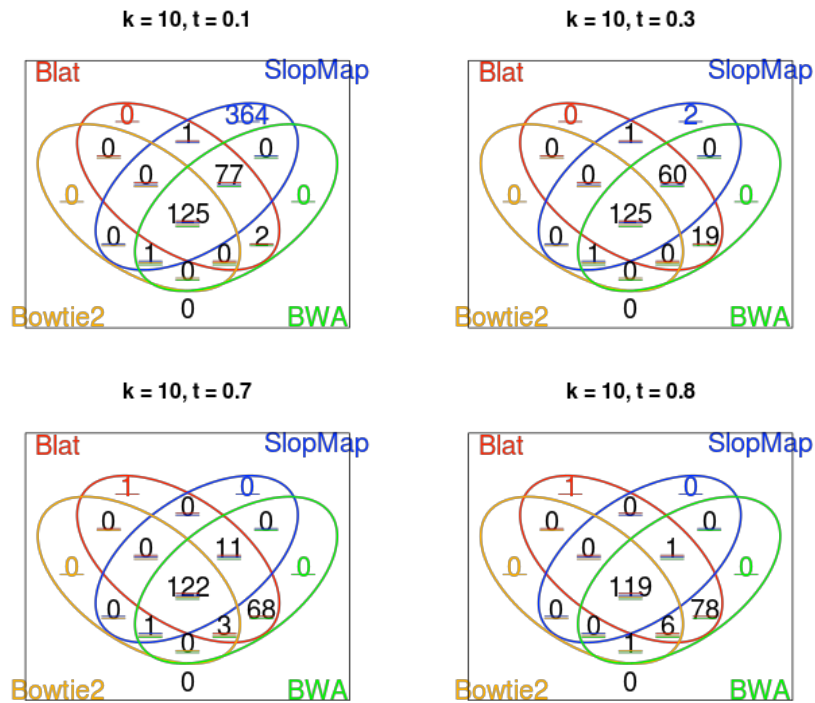

Figure 5: Number of reads reported by SlopMap and other tools depending on threshold values t and kmer sizes k = 10 bases. When t = 0.3 (the top-right figure), only 19 reads reported by Blat do not overlap with SlopMap, and those are also reported by BWA; reads reported by BWA and Bowtie2 are subsets of reads reported by SlopMap until t ≥ 0.7. These tests were counducted on Roche 454 E. coli K12 W583 DNA library used as a query set.

SlopMap is faster then Bowtie2 and BWA, but slower than Blat, which is the fastest of the mappers we tested. However, Blat requires that the input be in FastA format, and does not support writing output in FastA or FastQ format, making it necessary to perform additional steps, both before and after using the program. Post-mapping conversion work is also required for BWA and Bowtie2. Bowtie2 writes all sequences to the SAM file, and write unmapped or discordantly mapped reads to files using command line parameters to output mapped sequences in FastQ file format.

## 4.6 Memory requirements

SlopMap is fast and require very little memory (2-200MB, depending on target size). The memory consumption of SlopMap during searching grows linearly with the number of sequences and also depends on the kmer size defined by the user. For example, when the E. coli data set containing sequences with a mean length of 450 characters was indexed on 15-mers, 50 kB of memory was utilized for every 20,000 bases.

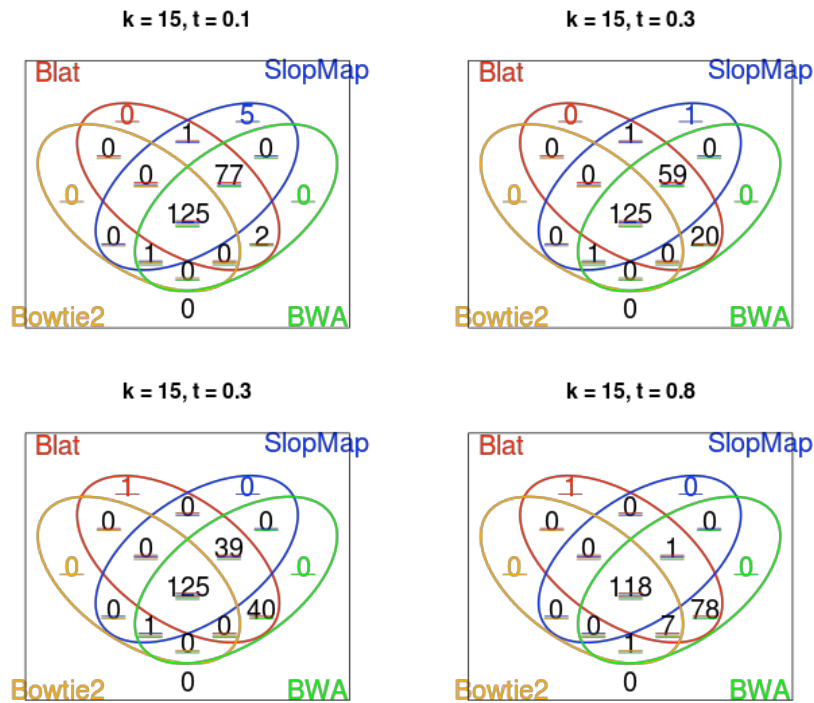

Figure 6: Number of reads reported by SlopMap and other tools depending on percent similarity threshold values t and kmer size k = 15 bases.

## 5 Conclusion and future work

SlopMap provides researchers with a high-throughput choice for searching large sets of reads against target sequences. The software presented is faster than some well-known aligners, sensitive to low-similarity matches when desired, and flexible enough to allow similarity comparison for DNA (and potentially RNA and proteins). SlopMap is specifically designed for matching queries against large (more than 500,000 sequences) query sets. Three of SlopMaps beneficial attributes are its speed, flexibility and ease of use. Despite being fast and efficient mapper, we plan to further improve SlopMap in the future by adding support for multicore execution and by exploring more space and time efficient methods for storing and looking up kmers.

We believe that the biological research community will benefit from using SlopMap.

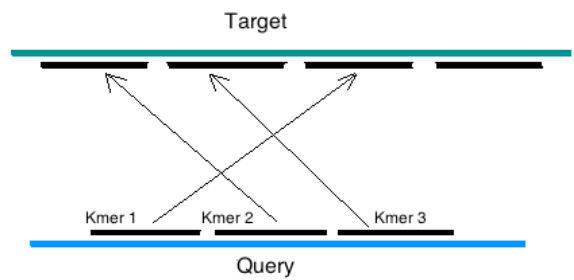

Figure 7: Non-consecutive kmer matching is still when using a target and query where rearrangements may have occurred. SlopMap identifies and reports these matches.

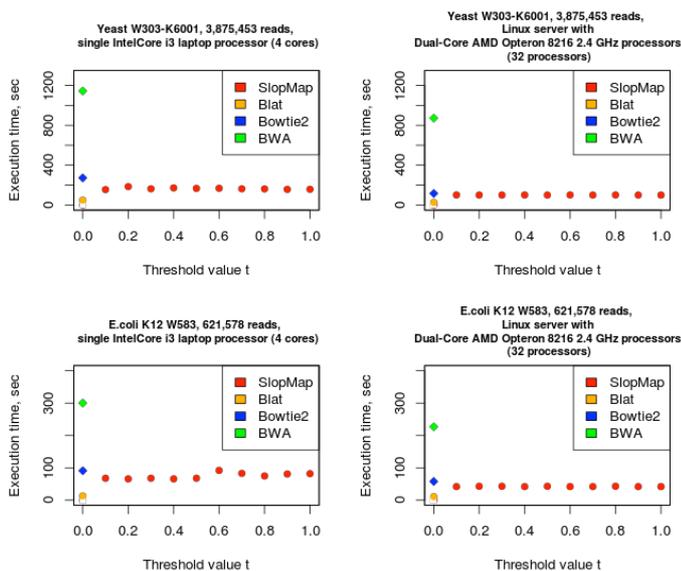

Figure 8: Search duration. Comparison of execution time between search tools. SlopMap is consistently fast across a range of threshold values (t), and perform somewhat faster than Bowtie2, and significantly faster than BWA while using the same amount of CPU resources.

# 6 Acknowledgments

This publication was made possible by NIH Grant P20 RR16454 from the IN-BRE Program of the National Center for Research Resources.